\documentclass[]{spie}  

\usepackage{amsmath,amsfonts,amssymb}
\usepackage{graphicx}
\usepackage[colorlinks=true, allcolors=blue]{hyperref}
\usepackage{multirow}

\newcommand{\keV}{\rm\thinspace keV}

\title{Identifying charged particle background events in X-ray imaging detectors with novel machine learning algorithms }

\author[a]{D.~R.~Wilkins}
\author[a]{S.~W.~Allen}
\author[b]{E.~D.~Miller}
\author[b]{M.~Bautz}
\author[a]{T.~Chattopadhyay}
\author[a]{S.~Fort}
\author[b]{C.~E.~Grant}
\author[a]{S.~Herrmann}
\author[c]{R.~Kraft}
\author[a]{R.~G.~Morris}
\author[c]{P.~Nulsen}
\affil[a]{Kavli Institute for Particle Astrophysics and Cosmology, Stanford University, 452 Lomita Mall, Stanford, CA 94305, USA}
\affil[b]{MIT Kavli Institute for Astrophysics and Space Research, 77 Massachusetts Avenue, Cambridge, MA 02139, USA}
\affil[c]{Harvard–Smithsonian Center for Astrophysics, 60 Garden Street, Cambridge, MA 02138, USA}

\authorinfo{Correspondence e-mail: dan.wilkins@stanford.edu}

\begin{document} 
\maketitle

\begin{abstract}
Space-based X-ray detectors are subject to significant fluxes of charged particles in orbit, notably energetic cosmic ray protons, contributing a significant background. We develop novel machine learning algorithms to detect charged particle events in next-generation X-ray CCDs and DEPFET detectors, with initial studies focusing on the \textit{Athena Wide Field Imager (WFI)} DEPFET detector. We train and test a prototype convolutional neural network algorithm  and find that charged particle and X-ray events are identified with a high degree of accuracy, exploiting correlations between pixels to improve performance over existing event detection algorithms. 99 per cent of frames containing a cosmic ray are identified and the neural network is able to correctly identify up to 40 per cent of the cosmic rays that are missed by current event classification criteria, showing potential to significantly reduce the instrumental background, and unlock the full scientific potential of future X-ray missions such as \textit{Athena}, \textit{Lynx} and \textit{AXIS}.
\end{abstract}

\keywords{X-ray astronomy, X-ray detector, X-ray satellite, background, CCD, DEPFET, machine learning, neural network}

\section{Introduction}
\label{sec:intro}  

Imaging detectors, based upon CCD (charged-coupled device) and similar technologies, have become the mainstay of space-based X-ray observatories. Pixelated detectors offer simultaneous imaging and spectroscopic capabilities, recording the spatial location, energy and time of individual photon events (assuming that the frame rate of the detector relative to the rate of incoming photons is such that a maximum of one photon is absorbed in each pixel per readout).

The Advanced CCD Imaging Spectrometer (ACIS) on the \textit{Chandra} X-ray observatory\cite{acis} has produced some of the highest spatial resolution images of the X-ray sky, enabling many scientific investigations, including the morphology of hot gas within clusters of galaxies (the intracluster medium, or ICM), the interactions of jets launched by supermassive black holes with their environments, resolved imaging of multiply-lensed quasars, and the detection of individual point sources (active galactic nuclei, or AGN) in deep-field X-ray surveys. While achieving lower spatial resolution than \textit{Chandra}, the enhanced collecting area of the \textit{XMM-Newton} X-ray observatory\cite{xmm} offers increased sensitivity to faint sources over the 0.3-10\keV\ energy range. Spectroscopy using the European Photon Imaging Camera (EPIC) cameras, and in particular the back-illuminated pn CCD\cite{xmm_epic}, has provided great insight into the close environments of black holes, and has enabled spectroscopic measurements of the temperature, density and metalicity of the ICM.

X-ray imaging detectors will continue to play a central role on the next generation X-ray observatories. In particular, the \textit{Athena} X-ray observatory\cite{athena}, scheduled for launch by the European Space Agency in the early 2030s, will offer an order of magnitude increase in collecting area over the current state-of-the-art. \textit{Athena} will carry the Wide-Field Imager (WFI)\cite{wfi}, a DEPFET detector, constructed from silicon and divided into pixels in a similar manner to a CCD. Alongside \textit{Athena}, the proposed flagship NASA X-ray observatory, \textit{Lynx}\cite{lynx}, would combine large collecting areas, comparable to \textit{Athena}, with exquisite spatial resolution imaging, comparable to \textit{Chandra}, provided by the \textit{High Definition X-ray Imager (HDXI)}\cite{lynx_hdxi}. On a smaller scale than the flagship X-ray observatories, the proposed probe-class \textit{Advanced X-ray Imaging Satellite (AXIS)}\cite{axis} would feature high angular resolution optics and will require a similar class of CCD-based imaging detector. While high-resolution spectroscopy on future X-ray observatories will largely be conducted using microcalorimeter arrays, such as the X-IFU on board \textit{Athena}, DEPFET and next-generation CCD detectors will still play a vital role in scientific investigations that simultaneously require a large field of view, good angular resolution and spectroscopic capability. Wide and deep surveys conducted with next-generation X-ray imagers aboard future observatories will yield precise measurements for vast samples of black holes, extending back to the epoch of cosmic dawn, to understand their formation and growth, while sensitive imaging of clusters and groups of galaxies, both nearby and at high redshift, will reveal the physics of the ICM and provide vital insight into the formation of large scale structure in the Universe.\cite{athena_science,lynx_science}

X-ray imaging detectors record signals not only in response to astrophysical X-rays that are received through the telescope, but also in response to charged particles. Charged particles producing signals in the detector include high energy cosmic ray protons (often referred to as `minimally ionizing particles' or MIPS) passing through the detector itself, or secondary protons, electrons and X-ray photons that are produced when charged particles interact with the spacecraft. Charged particles that impact X-ray satellites and produce components of the instrumental background arise from a number of sources: Galactic cosmic rays (GCRs), which include protons, electrons and helium ions with energies of tens of MeV to GeV; Solar energetic particles (SEP), which are mostly protons accelerated by the Sun to 10-100\,MeV; and protons accelerated in the heliosphere to hundreds of keV.\cite{wfi_bkg} In addition, low energy (`soft') protons of Solar origin, below 300\keV, can be deflected by the telescope's mirrors and focused onto the detector\cite{Fioretti1607.05319}.

When energy is deposited within the silicon detector by a photon or charged particle, a cloud of electrons is produced. This cloud diffuses outwards before reaching the readout gates resulting in the signal from a single event being spread across adjacent pixels.\cite{miller_charge_diff} Depending upon the size of the pixels and the location a photon is absorbed, a single X-ray photon can be manifested as a single, double or quadruple pixel event. A charged particle, however, depending on its trajectory, can produce signals in much larger groups of pixels, as energy is continually deposited as it passes through the silicon, and in multiple patches, as secondary particles produced by a proton interact separately with the silicon detector. 

In the current generation of event detection and reconstruction algorithms, as employed, for example, in the data reduction pipelines for \textit{Chandra} and \textit{XMM-Newton}, 
events are identified as isolated clusters of illuminated pixels in which signal is recorded above a threshold defined by the noise level in the pixels. The \texttt{PATTERN} or \texttt{GRADE} of the event is defined based upon the number of illuminated pixels and their arrangement, within what is usually a $3 \times 3$ grid of pixels (or a $5\times 5$ grid in the \textit{Chandra}`very faint source' mode) centered upon the pixel with the highest signal amplitude\cite{acis}. The total energy of the event (\textit{i.e.} the photon energy for an X-ray event) is computed by summing the signal amplitude in all of the illuminated pixels. A crude filter to exclude charged particle events is implemented by excluding events with total energy in excess of a photon that could have been focused by the telescope (the cut-off in the \textit{XMM-Newton} EPIC cameras is defined to be 15\keV), or by filtering based upon the \texttt{PATTERN} or \texttt{GRADE}, to exclude events spread over too many pixels to have been due to a single photon.

For satellites in relatively high orbits, the background signal induced by charged particle events can be significant, severely limiting the sensitivity of the detector to low surface brightness sources. 
Here, sources of interest include galaxy clusters, the largest gravitationally-bound structures in the Universe, and especially their outskirts, which are rich in astrophysical information.
\cite{Walker1810.00890} While simulations of cosmic ray interactions with the telescope and detector show that traditional event reconstruction and background filtering algorithms, based upon the total energy and number of adjacent pixels illuminated in an event, are able to remove $\sim$98 per cent of cosmic-ray induced background events \cite{wfi_bkg}, the remaining, unfiltered events still have a significant impact, severely limiting, for example, \textit{Chandra} and \textit{XMM-Newton} studies of observations of cluster outskirts and hampering studies of the formation and growth of the first supermassive black holes.

To fulfil the scientific potential of future X-ray missions such as \textit{Athena}, \textit{Lynx} and \textit{AXIS}, the ability to better understand and filter the instrumental background 
will be critical. We are exploring the ability of novel, artificial-intelligence (AI) event detection algorithms to do this. These algorithms identify X-ray and charged particle events in imaging X-ray detectors based not just upon the event energy and number of adjacent pixels illuminated, but on the morphology of events induced by charged particles and their secondaries across the entire detector plane. In Section~\ref{sec:geant4} we briefly discuss simulations of particle interactions with the spacecraft and detector upon which the algorithm development is based. In Section~\ref{sec:cnn} we describe a prototype classification algorithm, and in Section~\ref{sec:results} present the results of initial studies that demonstrate the feasibility of reducing the instrumental background with this new approach to event classification.

\section{Charged particle events in X-ray imaging detectors}
\label{sec:geant4}
Simulations of the interactions of X-ray photons and charged particles with a silicon DEPFET or CCD detector are central to understanding how each produces signals in the detector, and how we can more effectively detect and filter the instrumental background. Here, we consider the component of the background that is produced by Galactic cosmic rays, \textit{i.e.} primary protons. These protons may pass directly through the silicon detector and deposit energy among its pixels, or may interact with other parts of the spacecraft, producing secondary particles. These secondaries may be electrons generated in the ionization of the spacecraft material, or X-ray photons generated by fluorescence, bremsstrahlung, or inelastic scattering. In order to understand the signals induced in the silicon detector by the primary protons and their secondaries, and develop algorithms to identify and filter charged particle events, it is therefore necessary to model the interaction of the cosmic ray protons with both the spacecraft and detector.

\begin{figure} [ht]
\begin{center}
\begin{tabular}{c} 
\includegraphics[height=6cm]{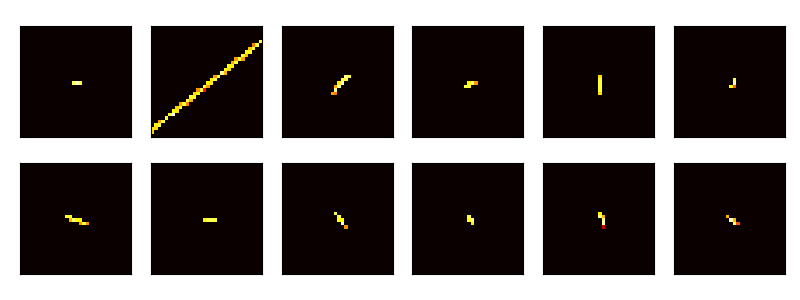}
\end{tabular}
\end{center}
\caption[gcr_events] 
{ \label{fig:gcr_events} 
Simulated \textsc{geant4} interactions of cosmic ray protons with the \textit{Athena WFI} X-ray detector. Protons may travel through the detector leaving long tracks of charge deposition. Alternatively, a proton may interact with a part of the spacecraft leading to a shower of secondary particles reaching the detector at once.}
\end{figure}

We base our study of charged particle events and background identification algorithms on simulations of the particle background conducted as part of the \textit{Athena Wide Field Imager} background study\cite{ou_geant4}\footnote{Simulations developed by the Open University (OU) and analyzed by the MIT group for the Athena Wide Field Imager Background Working Group}. The \textsc{geant4} code \cite{geant4} was used to trace cosmic ray protons, their secondaries and their interactions with the spacecraft and detector. \textsc{geant4} uses Monte Carlo methods to compute the passage of cosmic ray protons through the spacecraft. The simulation comprises a \textit{mass model} of the spacecraft with which particles may interact. The CCD or DEPFET detector itself is modelled as a sheet of silicon. As protons interact with material in the mass model, they deposit energy at each location and may produce one or more secondary particles (further protons, electrons and X-ray photons) that are additionally followed through the model, themselves depositing energy. The signal that would be recorded by the detector is generated by dividing the silicon element of the mass model into a grid of pixels, then summing the energy that is deposited in each pixel cell (notwithstanding the diffusion of charge, the voltage signal in each pixel corresponds to the deposited energy). The passage of each proton and its secondaries through the instrument is much faster than the integration time of a single detector image frame. We can therefore treat the the energy deposition from a single proton and its associated secondaries as occurring within the same detector frame. Simulations of the energy deposited per pixel as cosmic ray protons and their secondaries interact with the \textit{Athena WFI} DEPFET detector are shown in Figure~\ref{fig:gcr_events}.

These simulations can be compared to cosmic ray data gathered from a real CCD detector, using image frames that were taken when the filter wheel was in the closed position. Such a configuration blocks X-rays from reaching the detector such that all detected events must be due to cosmic rays. \textsc{geant4} simulations have been found to produce an accurate description of how cosmic ray protons interact with and are detected by the EPIC pn camera on board \textit{XMM-Newton} \cite{Bulbul1908.00604}.

\subsection{Simulated detector frames}
We simulate a set of 20,000 frames that would be read out from an X-ray imaging detector in order to train and test event classification algorithms. We consider small, $64\times64$ patches of a detector similar to the \textit{Athena WFI}, with $130\times130\,\mu\mathrm{m}$ pixels. Each frame contains a random combination of simulated cosmic ray induced charged particle events from the \textsc{geant4} simulation library.

We add to the \textsc{geant4} simulations of charged particle events a simplified description of astrophysical X-ray photons that reach the detector via the telescope mirrors. We randomly assign the location of each photon within the two-dimensional pixels and assume the energy is deposited at a single point. The diffusion of the electrons that are produced in response to this photon can then be simulated by placing a Gaussian function over this location, and signal is detected in pixels at which the Gaussian charge distribution is greater than the noise level of the detector. In the current generation of \textsc{geant4} simulations, the electrons do not diffuse from the locations of energy deposition, thus for consistency, we simulate X-ray events in which all of the signal is detected in a single pixel. The effects of charge diffusion will be explored in future work. In detectors such as the \textit{Athena WFI}, the frame rate is high enough (with at least one frame read out every 5\,ms) that for all but the brightest astrophysical point source, a maximum of one photon will received during each readout frame.

Each simulated frame may contain either one or two distinct events, which may be single cosmic ray or X-ray events, two cosmic ray events, or one cosmic ray event and one astrophysical photon, drawn at random. Each event is placed at a random location within the frame, at a random orientation. The final frame is then computed from the summed energy that was deposited into each pixel, representing the signal amplitudes that would be read out.

\section{Identifying X-ray and charged particle events with neural networks}
\label{sec:cnn}

We are developing a novel machine learning algorithm that will improve the accuracy of event classification and background filtering in imaging X-ray detectors, including DEPFET detectors such as the \textit{Athena Wide Field Imager} and next-generation CCDs, including the proposed \textit{Lynx HDXI}. The algorithm incorporates the detected signals in all telemetered pixels within a frame, rather than considering individual $3\times3$ clusters of pixels, to determine the optimal segmentation of each frame into individual events, and then identify the events as either X-rays or cosmic-ray induced background.

Such a holistic approach to frame segmentation and event classification has a number of advantages over traditional background filtering based upon the event energy and pixel pattern or grade. By considering patterns of charge deposition across all the pixels within a frame, nearby pixels that are illuminated following the interaction of a single proton with the spacecraft or detector can be considered as a single event, including the shower of secondary particles, which may produce their own events that are not contiguous with one another. Each event that is detected is assigned a probability of being a genuine astrophysical X-ray event, or an event due to a charged particle, and in the data analysis pipeline, events can be selected based upon a threshold probability value.

The observed cosmic ray charge patterns are governed by well-defined physical interactions that lead to specific predictions of the spread of the secondary particles and the observed correlation lengths between the illuminated pixels \cite{wfi_bkg}. In reality, however, these interactions are complex and probabilistic in nature such that it is not trivial to analytically derive criteria on which patterns can be filtered. A machine learning algorithm, however, is able to `learn' the rules that identify a charge pattern that is due to a cosmic ray interaction, as opposed to an X-ray, by observing a set of cases for which the answer is known. A machine learning algorithm is, for example, able to learn that low energy events that are due to secondary particles are associated with the primary proton track (while the traditional algorithm would only remove the track); or if there are multiple, nearby low energy events from secondaries produced as a proton interacts elsewhere on the spacecraft, that these are associated with one another, rather than being multiple, independent events that would previously have been identified as astrophysical X-rays.

\subsection{Development of a prototype frame classification algorithm}

We have developed a prototype machine learning algorithm that classifies an image (\textit{i.e.} the frame obtained in a single detector readout) as containing only astrophysical X-ray events, only cosmic ray events, or both astrophysical X-ray and cosmic ray events. The algorithm is based upon a convolutional neural network (CNN) and follows the architecture commonly employed in image recognition applications. The CNN forms an image recognition algorithm that classifies a frame (\textit{i.e.} the patterns of charge left in clusters of pixels by either X-ray or cosmic ray interactions) based upon features that are detected by a series of convolutional filters. Using convolutional filters for feature detection provides translational invariance; a given pattern will be classified in the same way wherever it appears within the image.

\begin{figure} [ht]
\begin{center}
\begin{tabular}{c} 
\includegraphics[height=15cm]{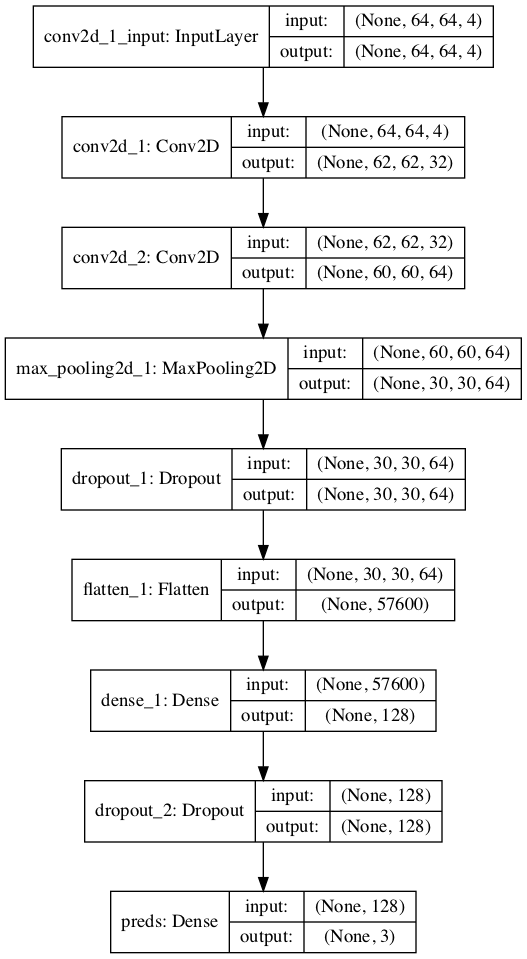}
\end{tabular}
\end{center}
\caption[cnn] 
{ \label{fig:cnn} 
Architecture of the convolutional neural network, constructed in \textsc{tensorflow} using \textsc{keras} to perform the classification of CCD and DEPFET frames. The left column shows the name and operation performed by each layer of the network, described in the text, while the right column shows the dimensions of the data input to and output from each layer. The dropout layers are a common component of neural networks, which break connections between some of the nodes within the layers, and are empirically found to reduce over-fitting.}
\end{figure}

The algorithm is constructed in the \textsc{tensorflow} framework\cite{tensorflow} and consists of two 2-dimensional convolutional layers that describe the features to be detected (each layer contains of a set of a $3\times3$ convolutional filters that slide over the input image), followed by ``max-pooling" layers that reduce the result of the convolutional filters applied to each patch of $3\times3$ pixels to a single summary value. In the prototype version of the algorithm, 32 filters or features are present in the first layer, and 64 in the second, although these numbers can be tuned to optimize the performance of the algorithm. A 128-feature fully connected (`dense') layer then classifies the frame based on the results of applying the convolutional filters, which is then connected to a 3-feature dense layer with `softmax' activation that yields the final classification of each frame. The three features of this final layer correspond to the three possible classifications of the frame; X-ray only, cosmic ray(s) only or both astrophysical X-rays and cosmic ray(s), and the activation is defined such that the values assigned to each of these classifications sums to unity. This means that the number assigned to each of these three classifications by the neural network can be interpreted as the `probability' that the frame fits into each classification. The model architecture is outlined in Figure~\ref{fig:cnn}.

Each of the convolutional filters is a matrix of free parameters, so too are the weightings of each input to the fully connected classification layers. These are termed \textit{hyperparameters}. The network is \textit{trained} by optimizing the values of the hyperparameters such that a \textit{training set} of images, for which the classification is known, are correctly classified. This is achieved by minimizing a \textit{loss function}, the \textit{binary cross-entropy}, that defines the classification errors for a given set of hyperparameter values. We construct the training set from a combination of the \textsc{geant4} simulations that show the energy deposited (and hence the signal recorded) in each pixel from a cosmic ray proton and its secondaries, and a sample of simulated X-ray events with different energies.  10,000 simulated frames (of which the contents are known) are used to train the network and fit the values of the hyperparameters.

In each frame, the pixel values correspond to the energy deposited in each pixel. As is common practice in image recognition with CNN algorithms, we normalize the image frames that are input to the neural network such that the maximum pixel value in each frame is 1. This allows the neural network to learn the shape of cosmic ray and X-ray events, rather than being able to directly associate the energy of specific pixels with the different events. The training set will necessarily be of finite size and not normalizing the input images can result in \textit{over-fitting} where the network focuses on overly-specific features of the training set that do not readily generalize to events beyond the training set. In order to maintain the information contained in the energy that is deposited in each pixel, which is an important discriminator between charged particles and X-ray photons, we divide each frame into energy channels. Each channel is itself an image frame, but containing only the pixels with values lying in defined energy ranges. The convolutional filters look for features in each of the separate channel images, as well as features between energy channels, in the same manner that CNNs are used to identify three-color RGB images. In the prototype algorithm, we split the images into three energy channels: pixels less than 5\keV, 5-10\keV\ and pixels above 10\keV, while also including the full frame image. The number of channels and the energy ranges of the channels can be tuned to optimize the performance of the algorithm.

Such an algorithm verifies the ability of a CNN to not only distinguish cosmic rays from X-rays, but to find an X-ray in the same frame as a cosmic ray and separate the events such that the cosmic ray can be discarded while maintaining the astrophysical signal. We define a cosmic ray event as any signal on the detector that is due to the interaction of a cosmic ray with the detector or spacecraft, whether that is the primary proton, secondary particles, or X-ray photons that are generated in the interaction. Astrophysical X-rays are the only X-ray photons that have reached the detector via the mirror. This definition enables the algorithm to associate cosmic-ray induced X-rays with nearby particles seen on the detector. Classifying regions of frames, and identifying whether one or more cosmic ray events is present, is the first step towards reducing the instrumental background and will enable regions of frames containing cosmic ray events to be excluded from the analysis.

\section{Results of feasibility studies}
\label{sec:results}

Once the prototype frame classification neural network algorithm had been trained, we verified its performance using a further set of 10,000 simulated frames, generated in the same manner as the training set. These test frames were not included in the training set and the network had not seen them before. We can therefore assess the ability of the algorithm to correctly classify the test frames and compare the fraction of cosmic ray events that are correctly identified with the fraction correctly identified by the traditional event classification method based on the energy and the number of illuminated pixels. The neural network yields a vector of three values for each frame, which can be interpreted as the probability that the frame fits into each of the three classes (containing X-rays only, cosmic rays only, or containing both X-rays and cosmic rays). The final classification of each frame is taken as that for which the assigned probability is the highest (although, if desired, more stringent criteria for the acceptance or rejection of events within a frame can be defined, requiring threshold values be reached in each class).

The results of these tests are summarized in Table~\ref{tab:results}. We find that the prototype CNN-based algorithm is highly successful identifying frames that contain cosmic-ray and X-ray events. We find that 99 per cent of all frames that contain a cosmic ray event of any sort (a proton track, electron and positron events or secondary X-ray photons) are  identified (\textit{i.e.} are classified as containing a cosmic ray only or both a cosmic ray and an X-ray). The false positive rate is very low --- a negligible number of clean frames containing only astrophysical X-ray photons are incorrectly classified as containing cosmic ray events (and would thus be incorrectly rejected). Of the frames that contained both X-ray and cosmic ray events, 97 per cent are correctly identified as containing both, while 3 per cent were identified as containing only a cosmic ray (for which the accompanying X-ray would be lost), demonstrating that in the majority of cases, X-rays can be distiguished from cosmic ray events within a single frame.

\begin{table}[ht]
\caption{Results of preliminary tests of the frame classification neural network, showing how simulated frames containing random combinations of astrophysical X-rays, cosmic rays and their secondaries produced during interactions with the spacecraft, and both X-rays and cosmic rays, were classified. For each frame, the full, raw, pixel data was input to the neural network.} 
\label{tab:results}
\begin{center}       
\begin{tabular}{|l|l|l|l|l|}
\hline
\rule[-1ex]{0pt}{3.5ex}  \multirow{2}{*}{\textbf{Input Frame}} & \multirow{2}{*}{\textbf{\# frames}} & \multicolumn{3}{c|}{\textbf{Number of frames identified as}} \\
\cline{3-5}
\rule[-1ex]{0pt}{3.5ex}  & & \textbf{X-ray only} & \textbf{Cosmic ray only} & \textbf{X-ray + cosmic ray}  \\
\hline
\rule[-1ex]{0pt}{3.5ex}  X-ray only & 2504 & 99.9\% & 0 & 0.1\%   \\
\hline
\rule[-1ex]{0pt}{3.5ex}  Cosmic rays only & 3724 & 1.2\% & 95.7\%& 3.1\%  \\
\hline
\rule[-1ex]{0pt}{3.5ex}  X-ray + cosmic ray & 3772 & 0.9\% & 2.5\% & 96.6\%  \\
\hline 
\end{tabular}
\end{center}
\end{table}

We may further assess the ability of the neural network to identify secondary particles produced when protons interact with the spacecraft (Table~\ref{tab:secondary}), including electrons and positrons, and X-ray photons (defining secondary photons to be part of cosmic ray events, distinct from astrophysical X-rays reaching the detector via the mirrors). We find that for a secondary X-ray photon accompanied by a charged particle produced by the same event on the detector, the algorithm is able to correctly identify the frame as containing only a cosmic ray event in 96 per cent of cases,  incorrectly identifying the frame as containing both a cosmic ray and astrophysical X-ray 4 per cent of the time. 
Current event filtering algorithms based upon energy and pixel pattern alone would not identify any of these secondary photons, since they are to all intents and purposes valid X-ray events. We also find that the algorithm is able to correctly identify 96 per cent of electron and positron events, though we caution that the number of such events in the simulation library is small. Electrons and positrons deposit energy in a single pixel and, in isolation, appear as valid X-ray events, though their coincidence with other electron/positron or secondary photon events enables them to be identified.

\begin{table}[ht]
\caption{Classification of frames by the neural network containing secondary photon events, accompanied by a charged particle, as well as frames containing only secondary electron and positron events, showing how the algorithm is able to use the coincidence between the secondaries and other particles resulting from the same event in a single frame to correctly identify the secondaries as cosmic ray events.} 
\label{tab:secondary}
\begin{center}       
\begin{tabular}{|l|l|l|l|l|}
\hline
\rule[-1ex]{0pt}{3.5ex}  \multirow{2}{*}{\textbf{Input Frame}} & \multirow{2}{*}{\textbf{\# frames}} & \multicolumn{3}{c|}{\textbf{Number of frames with secondaries identified as}} \\
\cline{3-5}
\rule[-1ex]{0pt}{3.5ex}  & & \textbf{X-ray only} & \textbf{Cosmic ray only} & \textbf{X-ray + cosmic ray}  \\
\hline
\rule[-1ex]{0pt}{3.5ex}  Photons + particles & 1038 & 0.1\% & 96.2\% & 3.7\%   \\
\hline
\rule[-1ex]{0pt}{3.5ex}  Electrons/positrons & 29 & 0 & 96.6\% & 3.4\%  \\
\hline 
\end{tabular}
\end{center}
\end{table}

\begin{figure} [ht]
\begin{center}
\begin{tabular}{c} 
\includegraphics[height=4cm]{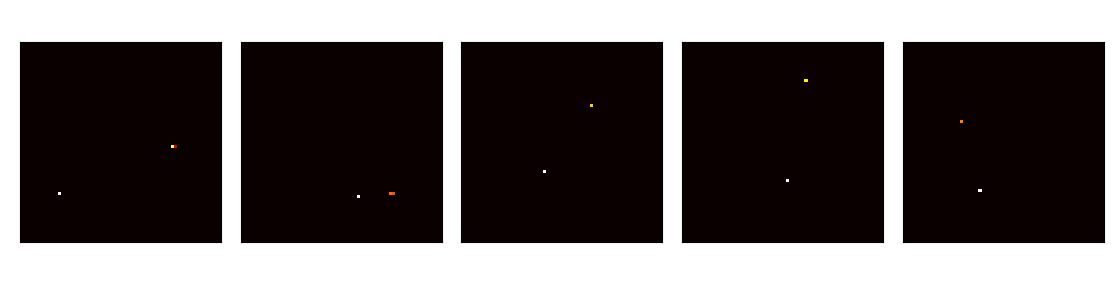}
\end{tabular}
\end{center}
\caption[undetect_gcr] 
{ \label{fig:undetect_gcr} 
Cosmic ray events that are undetected using existing filtering criteria based upon the event energy and \texttt{GRADE} or \texttt{PATTERN}, but are successfully identified by the neural network based upon the spatial correlations of multiple individual events within the same frame.}
\end{figure}

Of the 109 cosmic ray events that would not have been identified by the existing classification scheme using the event energy and pixel pattern, 39 per cent were correctly identified by our prototype CNN (Table~\ref{tab:results_undetect}). This includes frames that contain only undetected cosmic rays, or both an undetected cosmic ray event and an astrophysical X-ray, and the detection success rate is defined such that these frames are classified as containing a cosmic ray event, with or without an X-ray. Thus, in the case where all data are available from the detector and run through a simple CNN algorithm, we expect to achieve a 39 per cent reduction in the unrejected instrumental background compared with existing data analysis techniques. The gains of this simple CNN algorithm over traditional analysis approaches stem from its holistic approach to interpreting the frame. While with this prototype algorithm, each small, isolated group of illuminated pixels is not identified by itself, the appearance of multiple isolated groups and spatial correlations across the detector identifies the cosmic ray event, with the CNN recognizing that the probability of seeing multiple astrophysical X-ray events in the same frame is small when the frame rate is high. Figure~\ref{fig:undetect_gcr} shows examples of cosmic ray events missed by the standard event filtering scheme,  based upon the event energy and \texttt{GRADE} or \texttt{PATTERN}, that are successfully identified by the neural network.

\begin{table}[ht]
\caption{Results of the prototype frame classification algorithm identifying cosmic ray events that current event detection and classification criteria, based upon the total event energy and number of contiguous illuminated pixels, fail to identify.} 
\label{tab:results_undetect}
\begin{center}       
\begin{tabular}{|l|l|l|l|l|}
\hline
\rule[-1ex]{0pt}{3.5ex}  \multirow{2}{*}{\textbf{Input Frame}} & \multirow{2}{*}{\textbf{\# frames}} & \multicolumn{3}{c|}{\textbf{Number of undetected cosmic ray frames identified as}} \\
\cline{3-5}
\rule[-1ex]{0pt}{3.5ex}  & & \textbf{X-ray only} & \textbf{Cosmic ray only} & \textbf{X-ray + cosmic ray}  \\
\hline
\rule[-1ex]{0pt}{3.5ex}  Cosmic rays only & 53 & 71.7\% & 28.3\%& 0  \\
\hline
\rule[-1ex]{0pt}{3.5ex}  X-ray + cosmic ray & 56 & 51.3\% & 1.8\% & 46.4\%  \\
\hline 
\end{tabular}
\end{center}
\end{table}

\subsection{The completeness of charged particle data}
The above simulations assume that the full data generated in the detector by a charged particle event are available to the neural network. Sometimes early stages of event filtering are run on board X-ray astronomy satellites to reduce the volume of data telemetered to the ground. On board \textit{XMM-Newton}, data from the EPIC pn camera are subject to such filtering. When operated in the small window mode, all pixel data from all events (cosmic ray and X-ray) are telemetered. However, when the EPIC pn camera is operated in the large window, full frame or extended full frame modes that are typically used for the observation of extended X-ray sources, a degree of filtering takes place on board the spacecraft in order to limit the volume of data that is transferred to the ground. A simple filter is applied on-board to remove the majority of cosmic ray events: if a column of the detector contains any pixel above the 15\,keV threshold defined for cosmic ray detection, all pixels from that column, and the column either side of it, are discarded. While this filtering removes a significant number of cosmic ray events from the observations, in particular the tracks left as protons traverse the detector, it leaves behind the smaller, low energy events produced by the secondary particles from interactions elsewhere in the spacecraft, and the secondary events that can branch off major proton tracks. The residual background event rate is significant in observations of low surface brightness X-ray sources. Because these data are discarded on board the satellite and not available in the archive, some of the information that may associate the smaller secondary events with larger proton tracks are not available to our algorithm. For example, the telemetered data do not indentify the specific columns that were discarded during the frame, although the total number of columns discarded from each frame is available in the recorded data.

In order to test the performance of the CNN when such pre-filtering has been run on the input data, we retrain the network using only the filtered event data. We find that the overall accuracy of the neural network drops (Table~\ref{tab:results_mipfilter}). The remaining, unfiltered cosmic ray events are similar in appearance to X-ray events. Where previously, in the case of no on-board filtering, small, low energy particle events had been identified by association with larger particle tracks in the same frame, the information about these tracks has been removed from the frame data. Resultingly, only 63.5 per cent of all frames containing a cosmic ray event are correctly identified as such. Compared to current algorithms, the performance of the prototype network is still impressive, however,  
with $80$ per cent of the cosmic ray events missed by standard event energy and pattern criteria being correctly identified in the pre-filtered frames, although this comes at the expense of a 21 per cent false positive rate, \textit{i.e.} 21 per cent of frames containing only genuine astrophysical X-ray events are incorrectly identified as containing a cosmic ray event. We conclude that it is important for the full pixel data from each detector frame to be available to the neural network, so that cosmic ray events can be accurately identified without removing genuine astrophysical X-rays.

\begin{table}[ht]
\caption{Results of the prototype frame classification algorithm when the pixel data have been pre-filtered following the on-board filtering prescription employed by the \textit{XMM-Newton} EPIC pn camera.} 
\label{tab:results_mipfilter}
\begin{center}       
\begin{tabular}{|l|l|l|l|l|}
\hline
\rule[-1ex]{0pt}{3.5ex}  \multirow{2}{*}{\textbf{Input Frame}} & \multirow{2}{*}{\textbf{\# frames}} & \multicolumn{3}{c|}{\textbf{Number of pre-filtered frames identified as}} \\
\cline{3-5}
\rule[-1ex]{0pt}{3.5ex}  & & \textbf{X-ray only} & \textbf{Cosmic ray only} & \textbf{X-ray + cosmic ray}  \\
\hline
\rule[-1ex]{0pt}{3.5ex}  X-ray only & 2645 & 79.5\% & 20.5 & 0\%   \\
\hline
\rule[-1ex]{0pt}{3.5ex}  Cosmic rays only & 3675 & 36.4\% & 49.6\%& 14.0\%  \\
\hline
\rule[-1ex]{0pt}{3.5ex}  X-ray + cosmic ray & 3680 & 1.3\% & 4.8\% & 93.9\%  \\
\hline 
\end{tabular}
\end{center}
\end{table}

\section{From frame classification to event classification}
For a large X-ray imaging detector, the probability of any given frame containing a cosmic ray event is near unity. Therefore, in order to preserve the astrophysical signal, a frame cannot simply be discarded; the X-ray and cosmic ray events must be separated. An image may be a frame read out from the entire detector chip, or a smaller region of that frame in which a discrete group of events is seen. Performing filtering on smaller regions of the frame will allow the same algorithm to remove the cosmic ray events (since \textsc{geant4} simulations show them to be localized) while retaining almost all of the X-rays. To preserve the X-rays that appear close to the cosmic ray events, however, it is necessary to classify the individual events, rather than entire frames or sub-frames.

Once the algorithm has been trained to successfully classify frames and regions of frames, the next stage is to extend it to perform \textit{image segmentation}, that is the detection and classification of an arbitrary number of individual events within a frame, rather than flagging the whole frame or region of the frame for exclusion or inclusion in the analysis. An image segmentation algorithm can be constructed from a convolutional neural network following standard approaches in computer vision. This is commonly achieved by adding further layers to the end of the network, which, following the feature detection in the early layers, `up-sample' the results to identify and classify features in separate parts of the image, either classifying individual pixels, or clusters of neighboring pixels.\cite{unet}

Such an image segmentation algorithm is trained to optimally group the pixels into individual events and then assign to each of the detected event a classification that represents the probability that it is due to an astrophysical X-ray photon or a cosmic ray. Event filtering will be conducted by defining a threshold value; if the cosmic ray probability is above the threshold, the event may excluded from the analysis of the X-ray data. The full image segmentation algorithm will be presented in a future work.

\begin{figure} [ht]
\begin{center}
\begin{tabular}{c} 
\includegraphics[height=7cm]{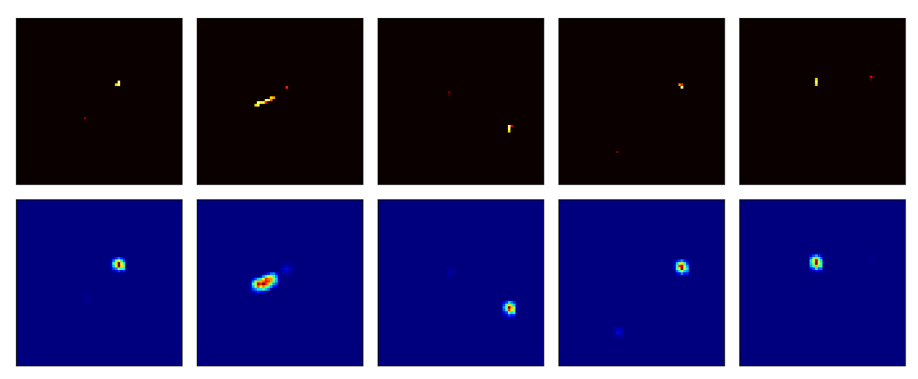}
\end{tabular}
\end{center}
\caption[saliency] 
{ \label{fig:saliency} 
\textit{Top row:} simulated \textit{Athena WFI} frames containing both an X-ray and cosmic ray event. \textit{Bottom row:} Saliency maps corresponding to each frame, showing the derivative of the `decision' of the algorithm with respect to each pixel. We see how the network activates on the pixels illuminated by the cosmic rays, and not those illuminated by X-rays, leading to the identification of the cosmic ray event by the algorithm.}
\end{figure}

We can demonstrate the capability of the prototype frame classification algorithm to identify features attributable to cosmic ray \textit{vs.} X-ray events by constructing a \textit{saliency map} from the frame classification neural network. The saliency map is computed from the derivative of the output classification with respect to the value of each pixel, highlighting the pixels in the image that caused the neural network to make the `decision' that it did. Figure~\ref{fig:saliency} shows a sample of frames containing both cosmic ray and X-ray events, along with their saliency maps with respect to the `cosmic ray' classification. It can be seen that in each case, the network is correctly identifying the pixels illuminated by the cosmic ray, which show significantly higher saliency values than the pixels illuminated by the X-ray.

\section{Conclusions}
\label{sec:conclusions}
We have demonstrated the feasibility of employing machine learning algorithms based on neural networks to identify charged particle events, due to cosmic rays, in X-ray imaging detectors (including DEPFET and next-generation CCD detectors), and to separate this component of the instrumental background from the astrophysical X-rays that are sought.

A prototype algorithm, based upon a convolutional neural network (CNN), is able to classify individual frames read out from a DEPFET or CCD detector as containing only genuine, astrophysical X-ray events, only cosmic-ray induced charged particle events, or both X-ray and particle events. The prototype algorithm performs with a high degree of accuracy, successfully identifying 99 per cent of frames containing a cosmic ray. The false positive rate is very low, and only 2.5 per cent of frames containing genuine X-ray events are incorrectly classified as containing a cosmic ray.

The neural network algorithm is able to correctly identify up to 40 per cent of the cosmic ray events that are missed by current event classification criteria. Employing artificial intelligence in the analysis of the raw, pixel-level data from next-generation X-ray CCDs and DEPFETs therefore holds the potential to significantly reduce the instrumental background.

When early stage filtering of the raw CCD frame data is performed on board the spacecraft, the neural network can be specifically trained on events that are missed by traditional event filtering. In this case, up to 80 per cent of cosmic ray induced charged particle events can be identified, though at the expense of a high false positive rate of 22 per cent. These findings underscore the importance of having the full set of data from charged particle events available to the algorithm to maximize performance.

Following the successful development of neural network frame classification algorithms, image segmentation algorithms can be implemented that take a holistic approach to event detection in next-generation X-ray imaging detectors. Considering the data from all pixels together, the algorithm will optimally segment each frame into individual events and determine the probability of each being due to a cosmic ray. Such an approach shows potential to significantly reduce the instrumental background, and unlock the full scientific potential of future X-ray missions such as \textit{Athena}, \textit{Lynx} and \textit{AXIS}.

\acknowledgments 
 
We thank Jonathan Keelan for providing \textsc{geant4} simulations of particle interactions with the \textit{Athena WFI} detector, in addition to routines and guidance for analyzing the \textsc{geant4} output. This work has been supported by the US \textit{Athena Wide Field Imager} Instrument Consortium under NASA grant NNX17AB07G and by the U.S. Department of Energy under contract number DE-AC02-76SF00515. DRW received additional support for the duration of this work under Einstein Postdoctoral Fellowship grant number PF6-170160, awarded by the Chandra X-ray Center, operated by the Smithsonian Astrophysical Observatory for NASA under contract NAS8-03060, and from a Kavli Fellowship at Stanford University.

\bibliography{ai_bkg} 

\begin{thebibliography}{10}

\bibitem{acis}
{Garmire}, G.~P., {Bautz}, M.~W., {Ford}, P.~G., {Nousek}, J.~A., and {Ricker},
  George~R., J., ``{Advanced CCD imaging spectrometer (ACIS) instrument on the
  Chandra X-ray Observatory},'' in [{\em X-Ray and Gamma-Ray Telescopes and
  Instruments for Astronomy.}{\nolinebreak\hspace{0.1em}]},  {Truemper}, J.~E.
  and {Tananbaum}, H.~D., eds., {\em Society of Photo-Optical Instrumentation
  Engineers (SPIE) Conference Series} {\bf 4851},  28--44 (Mar. 2003).

\bibitem{xmm}
{Jansen}, F., {Lumb}, D., {Altieri}, B., {Clavel}, J., {Ehle}, M., {Erd}, C.,
  {Gabriel}, C., {Guainazzi}, M., {Gondoin}, P., {Much}, R., {Munoz}, R.,
  {Santos}, M., {Schartel}, N., {Texier}, D., and {Vacanti}, G., ``{XMM-Newton
  observatory. I. The spacecraft and operations},'' {\em \aap}~{\bf 365},
  L1--L6 (Jan. 2001).

\bibitem{xmm_epic}
{Str{\"u}der}, L., {Briel}, U., {Dennerl}, K., {Hartmann}, R., {Kendziorra},
  E., {Meidinger}, N., {Pfeffermann}, E., {Reppin}, C., {Aschenbach}, B.,
  {Bornemann}, W., {Br{\"a}uninger}, H., {Burkert}, W., and {Elender}, M.,
  ``{The European Photon Imaging Camera on XMM-Newton: The pn-CCD camera},''
  {\em \aap}~{\bf 365},  L18--L26 (Jan. 2001).

\bibitem{athena}
Barcons, X., Barret, D., Decourchelle, A., den Herder, J.~W., Fabian, A.~C.,
  Matsumoto, H., Lumb, D., Nandra, K., Piro, L., Smith, R.~K., and Willingale,
  R., ``Athena: Esa's x-ray observatory for the late 2020s,'' {\em
  Astronomische Nachrichten}~{\bf 338}(2-3),  153--158 (2017).

\bibitem{wfi}
Meidinger, N., Nandra, K., and Plattner, M., ``{Development of the Wide Field
  Imager instrument for ATHENA},'' in [{\em Space Telescopes and
  Instrumentation 2018: Ultraviolet to Gamma Ray}{\nolinebreak\hspace{0.1em}]},
   den Herder, J.-W.~A., Nikzad, S., and Nakazawa, K., eds.,  {\bf 10699},  312
  -- 323, International Society for Optics and Photonics, SPIE (2018).

\bibitem{lynx}
{Gaskin}, J.~A., {Swartz}, D.~A., {Vikhlinin}, A., {{\"O}zel}, F., {Gelmis},
  K.~E., {Arenberg}, J.~W., {Bandler}, S.~R., {Bautz}, M.~W., {Civitani},
  M.~M., {Dominguez}, A., {Eckart}, M.~E., {Falcone}, A.~D.,
  {Figueroa-Feliciano}, E., {Freeman}, M.~D., {G{\"u}nther}, H.~M., {Havey},
  K.~A., {Heilmann}, R.~K., {Kilaru}, K., {Kraft}, R.~P., {McCarley}, K.~S.,
  {McEntaffer}, R.~L., {Pareschi}, G., {Purcell}, W., {Reid}, P.~B.,
  {Schattenburg}, M.~L., {Schwartz}, D.~A., {Schwartz}, E.~D., {Tananbaum},
  H.~D., {Tremblay}, G.~R., {Zhang}, W.~W., and {Zuhone}, J.~A., ``{Lynx X-Ray
  Observatory: an overview},'' {\em Journal of Astronomical Telescopes,
  Instruments, and Systems}~{\bf 5},  021001 (Apr. 2019).

\bibitem{lynx_hdxi}
{Falcone}, A.~D., {Kraft}, R.~P., {Bautz}, M.~W., {Gaskin}, J.~A., {Mulqueen},
  J.~A., and {Swartz}, D.~A., ``{The high definition x-ray imager (HDXI)
  instrument on the Lynx X-ray Surveyor},'' in [{\em Space Telescopes and
  Instrumentation 2018: Ultraviolet to Gamma Ray}{\nolinebreak\hspace{0.1em}]},
   {den Herder}, J.-W.~A., {Nikzad}, S., and {Nakazawa}, K., eds., {\em Society
  of Photo-Optical Instrumentation Engineers (SPIE) Conference Series} {\bf
  10699},  1069912 (July 2018).

\bibitem{axis}
{Mushotzky}, R., ``{AXIS: a probe class next generation high angular resolution
  x-ray imaging satellite},'' in [{\em Space Telescopes and Instrumentation
  2018: Ultraviolet to Gamma Ray}{\nolinebreak\hspace{0.1em}]},  {den Herder},
  J.-W.~A., {Nikzad}, S., and {Nakazawa}, K., eds., {\em Society of
  Photo-Optical Instrumentation Engineers (SPIE) Conference Series} {\bf
  10699},  1069929 (July 2018).

\bibitem{athena_science}
{Rau}, A., {Nandra}, K., {Aird}, J., {Comastri}, A., {Dauser}, T., {Merloni},
  A., {Pratt}, G.~W., {Reiprich}, T.~H., {Fabian}, A.~C., {Georgakakis}, A.,
  {G{\"u}del}, M., {R{\'o}{\.Z}a{\'n}ska}, A., {Sanders}, J.~S., {Sasaki}, M.,
  {Vaughan}, S., {Wilms}, J., and {Meidinger}, N., ``{Athena Wide Field Imager
  key science drivers},'' in [{\em Space Telescopes and Instrumentation 2016:
  Ultraviolet to Gamma Ray}{\nolinebreak\hspace{0.1em}]},  {den Herder},
  J.-W.~A., {Takahashi}, T., and {Bautz}, M., eds., {\em Society of
  Photo-Optical Instrumentation Engineers (SPIE) Conference Series} {\bf 9905},
   99052B (July 2016).

\bibitem{lynx_science}
{Bautz}, M.~W., ``{The Lynx X-Ray Observatory: Science Drivers},'' in [{\em UV,
  X-Ray, and Gamma-Ray Space Instrumentation for Astronomy
  XXI}{\nolinebreak\hspace{0.1em}]},  {\em Society of Photo-Optical
  Instrumentation Engineers (SPIE) Conference Series} {\bf 11118},  111180J
  (Sept. 2019).

\bibitem{wfi_bkg}
{von Kienlin}, A., {Eraerds}, T., {Bulbul}, E., {Fioretti}, V., {Gastaldello},
  F., {Grant}, C.~E., {Hall}, D., {Holland}, A., {Keelan}, J., {Meidinger}, N.,
  {Molendi}, S., {Perinati}, E., and {Rau}, A., ``{Evaluation of the ATHENA/WFI
  instrumental background},'' in [{\em \procspie}{\nolinebreak\hspace{0.1em}]},
   {\em Society of Photo-Optical Instrumentation Engineers (SPIE) Conference
  Series} {\bf 10699},  106991I (July 2018).

\bibitem{Fioretti1607.05319}
{Fioretti}, V., {Bulgarelli}, A., {Malaguti}, G., {Spiga}, D., and {Tiengo},
  A., ``{Monte Carlo simulations of soft proton flares: testing the physics
  with XMM-Newton},'' in [{\em \procspie}{\nolinebreak\hspace{0.1em}]},  {\em
  Society of Photo-Optical Instrumentation Engineers (SPIE) Conference Series}
  {\bf 9905},  99056W (July 2016).

\bibitem{miller_charge_diff}
{Miller}, E.~D., {Foster}, R., {Lage}, C., {Prigozhin}, G., {Bautz}, M.,
  {Grant}, C., {LaMarr}, B., and {Malonis}, A., ``{The effects of charge
  diffusion on soft x-ray response for future high-resolution imagers},'' in
  [{\em Space Telescopes and Instrumentation 2018: Ultraviolet to Gamma
  Ray}{\nolinebreak\hspace{0.1em}]},  {den Herder}, J.-W.~A., {Nikzad}, S., and
  {Nakazawa}, K., eds., {\em Society of Photo-Optical Instrumentation Engineers
  (SPIE) Conference Series} {\bf 10699},  106995R (July 2018).

\bibitem{Walker1810.00890}
{Walker}, S., {Simionescu}, A., {Nagai}, D., {Okabe}, N., {Eckert}, D.,
  {Mroczkowski}, T., {Akamatsu}, H., {Ettori}, S., and {Ghirardini}, V., ``{The
  Physics of Galaxy Cluster Outskirts},'' {\em \ssr}~{\bf 215},  7 (Jan. 2019).

\bibitem{ou_geant4}
Hall, D., Keelan, J., Davis, C., Hetherington, O., Leese, M., and Holland, A.,
  ``{Predicting the particle-induced background for future x-ray astronomy
  missions: the importance of experimental validation for GEANT4
  simulations},'' in [{\em High Energy, Optical, and Infrared Detectors for
  Astronomy VIII}{\nolinebreak\hspace{0.1em}]},  Holland, A.~D. and Beletic,
  J., eds.,  {\bf 10709},  762 -- 772, International Society for Optics and
  Photonics, SPIE (2018).

\bibitem{geant4}
Agostinelli, S., Allison, J., Amako, K., Apostolakis, J., Araujo, H., Arce, P.,
  Asai, M., Axen, D., Banerjee, S., Barrand, G., Behner, F., Bellagamba, L.,
  Boudreau, J., Broglia, L., Brunengo, A., Burkhardt, H., Chauvie, S., Chuma,
  J., Chytracek, R., Cooperman, G., Cosmo, G., Degtyarenko, P., Dell'Acqua, A.,
  Depaola, G., Dietrich, D., Enami, R., Feliciello, A., Ferguson, C.,
  Fesefeldt, H., Folger, G., Foppiano, F., Forti, A., Garelli, S., Giani, S.,
  Giannitrapani, R., Gibin, D., Cadenas], J.~G., González, I., Abril], G.~G.,
  Greeniaus, G., Greiner, W., Grichine, V., Grossheim, A., Guatelli, S.,
  Gumplinger, P., Hamatsu, R., Hashimoto, K., Hasui, H., Heikkinen, A., Howard,
  A., Ivanchenko, V., Johnson, A., Jones, F., Kallenbach, J., Kanaya, N.,
  Kawabata, M., Kawabata, Y., Kawaguti, M., Kelner, S., Kent, P., Kimura, A.,
  Kodama, T., Kokoulin, R., Kossov, M., Kurashige, H., Lamanna, E., Lampén,
  T., Lara, V., Lefebure, V., Lei, F., Liendl, M., Lockman, W., Longo, F.,
  Magni, S., Maire, M., Medernach, E., Minamimoto, K., de~Freitas], P.~M.,
  Morita, Y., Murakami, K., Nagamatu, M., Nartallo, R., Nieminen, P.,
  Nishimura, T., Ohtsubo, K., Okamura, M., O'Neale, S., Oohata, Y., Paech, K.,
  Perl, J., Pfeiffer, A., Pia, M., Ranjard, F., Rybin, A., Sadilov, S., Salvo],
  E.~D., Santin, G., Sasaki, T., Savvas, N., Sawada, Y., Scherer, S., Sei, S.,
  Sirotenko, V., Smith, D., Starkov, N., Stoecker, H., Sulkimo, J., Takahata,
  M., Tanaka, S., Tcherniaev, E., Tehrani], E.~S., Tropeano, M., Truscott, P.,
  Uno, H., Urban, L., Urban, P., Verderi, M., Walkden, A., Wander, W., Weber,
  H., Wellisch, J., Wenaus, T., Williams, D., Wright, D., Yamada, T., Yoshida,
  H., and Zschiesche, D., ``Geant4—a simulation toolkit,'' {\em Nuclear
  Instruments and Methods in Physics Research Section A: Accelerators,
  Spectrometers, Detectors and Associated Equipment}~{\bf 506}(3),  250 -- 303
  (2003).

\bibitem{Bulbul1908.00604}
{Bulbul}, E., {Kraft}, R., {Nulsen}, P., {Freyberg}, M., {Miller}, E.~D.,
  {Grant}, C., {Bautz}, M.~W., {Burrows}, D.~N., {Allen}, S., {Eraerds}, T.,
  {Fioretti}, V., {Gastaldello}, F., {Ghirardini}, V., {Hall}, D., {Meidinger},
  N., {Molendi}, S., {Rau}, A., {Wilkins}, D., and {Wilms}, J.,
  ``{Characterization of the Particle-induced Background of XMM-Newton EPIC-pn:
  Short- and Long-term Variability},'' {\em \apj}~{\bf 891},  13 (Mar. 2020).

\bibitem{tensorflow}
Abadi, M., Agarwal, A., Barham, P., Brevdo, E., Chen, Z., Citro, C., Corrado,
  G.~S., Davis, A., Dean, J., Devin, M., Ghemawat, S., Goodfellow, I., Harp,
  A., Irving, G., Isard, M., Jia, Y., Jozefowicz, R., Kaiser, L., Kudlur, M.,
  Levenberg, J., Man\'{e}, D., Monga, R., Moore, S., Murray, D., Olah, C.,
  Schuster, M., Shlens, J., Steiner, B., Sutskever, I., Talwar, K., Tucker, P.,
  Vanhoucke, V., Vasudevan, V., Vi\'{e}gas, F., Vinyals, O., Warden, P.,
  Wattenberg, M., Wicke, M., Yu, Y., and Zheng, X., ``{TensorFlow}: Large-scale
  machine learning on heterogeneous systems,'' (2015).
\newblock Software available from tensorflow.org.

\bibitem{unet}
Ronneberger, O., Fischer, P., and Brox, T., ``U-net: Convolutional networks for
  biomedical image segmentation,'' (2015).

\end{thebibliography}
\bibliographystyle{spiebib} 

\end{document}